\documentclass[superscriptaddress,reprint,amsmath,amssymb,aps,floatfix]{revtex4-1}

\usepackage{amsmath,gensymb,textcomp,bm,dcolumn,eurosym,array,tabu,multirow,nicefrac,color,subfigure,graphicx,upgreek}
\usepackage[colorlinks, linkcolor=blue, citecolor=blue, urlcolor=blue, breaklinks=true]{hyperref}

\usepackage{mathpazo,times} 

\newcommand{\ket}[1]{\lvert #1 \rangle}

\usepackage[subsectionbib]{bibunits}
\usepackage{graphicx}
\usepackage{dcolumn}
\usepackage{bm,MnSymbol}
\usepackage{units}
\usepackage{psfrag}
\usepackage{float}
\usepackage{color}

\begin{document}

\preprint{APS/123-QED}

\title{Harnessing high-dimensional symmetric and anti-symmetric Bell states through quantum interference}

\author{Ling Hong}
 \affiliation{Department of Physics, Xiamen University, Xiamen 361005, China}
\author{Yuning Zhang}
 \affiliation{Department of Physics, Xiamen University, Xiamen 361005, China}
\author{Yuanyuan Chen}
 \email{chenyy@xmu.edu.cn}
 \affiliation{Department of Physics, Xiamen University, Xiamen 361005, China}
\author{Lixiang Chen}%
 \email{chenlx@xmu.edu.cn}
\affiliation{Department of Physics, Xiamen University, Xiamen 361005, China}

\begin{abstract}
High-dimensional quantum entanglement is an essential resource in quantum technology since it provides benefits in increasing the information capacity and processing speed. Thus, the controlled harnessing of high-dimensional entanglement has long been hailed as a necessary prerequisite towards practical quantum applications. By using a deterministic quantum state filter that implemented through quantum interference, we present a generalised formulation for the complete high-dimensional symmetric and anti-symmetric Bell basis, and experimentally prepare four-dimensional orbital angular momentum Bell states that provide the well-behaved symmetric or anti-symmetric properties. Additionally, we use a concise yet efficient scan of temporal delay to directly observe high-dimensional two-photon interference effects in spatial modes. These results provide an alternative way for harnessing high-dimensional entanglement, and may facilitate the use of quantum interference for more complex quantum information processing tasks that beyond qubits.
\end{abstract}

\maketitle


\section{Introduction}
It has been the broad consensus that high-dimensional entanglement has various benefits in quantum information technologies, such as increasing the security and channel capacity in communication systems, and even speeding up certain tasks in photonic quantum computation \cite{ecker2019overcoming,cozzolino2019high,erhard2020advances,raussendorf2007fault,zhang2016experimental}. While there are several physical attributes can be used to realize high-dimensional quantum states, orbital angular momentum (OAM) provides access to the inherent infinite-dimensional state space and the flexible engineering, and thus it has been investigated in a wide range of applications including communication \cite{liu2020orbital,fang2020orbital}, imaging and sensing \cite{ren2019metasurface,courtial1998measurement,lavery2013detection,zhao2023vortex}, to name but a few. As a good candidate for large-scale entangled quantum states, harnessing the OAM entanglement still faces many technical challenges. For example, the complete high-dimensional Bell basis forms the foundation for many quantum protocols like superdense coding and quantum teleportation \cite{barreiro2008beating,pirandola2015advances,ren2017ground}. Although the preparation of these Bell states can be implemented by operating the well-designed cyclic transformation or adaptive pump modulation \cite{wang2017generation,chen2020coherent}, their experimental performances are strictly limited by the increase of entanglement dimensionality.

Hong-Ou-Mandel (HOM) interference states the fact that identical photons that arrive simultaneously on different input ports of a balanced beam splitter would bunch into a common output port \cite{hong1987measurement,liu2022hong}. This quantum phenomenon enables a large number of quantum information processing tasks including the measurement of optical delays between different paths \cite{chen2019hong,lyons2018attosecond}, the characterization of single photon emitters \cite{thoma2016exploring,ollivier2021hong}, and the implementation of Bell state measurement for quantum teleportation and entanglement swapping \cite{williams2017superdense,hu2020experimental,luo2019quantum,liu2022all}. Analogously, the HOM interference of high-dimensional entanglement is a necessary prerequisite towards high-dimensional quantum applications. To tackle this issue, Zhang \textit{et al.} presented a well-designed scheme for the controlled engineering of two-photon high-dimensional states entangled in their OAM through quantum interference, and demonstrate an elaborate approach to implement simultaneous entanglement swapping of multiple OAM states of light \cite{zhang2016engineering,zhang2017simultaneous}. While two-photon interference in multiple transverse-spatial modes along a single beam-path is explored by using the unitary transformation implemented by the multiplane light conversion technic, the lossy method of generating the spatial modes and the limited efficiency of spatial light modulators limit the observation of HOM interference for higher-dimensional and variously-formed quantum entanglement \cite{hikkam2021high}. \textcolor{black}{More recently, several elaborate experimental implementations for exploring the four- and eight-dimensional symmetric and anti-symmetric Bell state based on hyperentanglement between polarization and OAM modes have already been proposed, which are applied in observing particle exchange phase and enhancing the channel capacity of quantum protocol \cite{kong2019manipulation,liu2022hong}. However, the engineering of the complete high-dimensional symmetric and anti-symmetric Bell states and the observation of their quantum interference in a single degree of freedom has been explored relatively little.}

\textcolor{black}{A general approach for constructing a basis of maximally entangled multidimensional bipartite states has already been presented \cite{sych2009a}.} For example, the conventional formulation of d-dimensional Bell states of a bipartite system can be expressed as $\ket{\phi}_{m,n}=\frac{1}{\sqrt{d}}\sum_{k=0}^{d-1}e^{i2\pi nk/d}\ket{k}_A\ket{m\oplus k}_B$, where $m\oplus k=(m+k)$ $mod$ $d$, the variable $m$ defines the OAM correlation between paired photons, and the variable $n$ defines the relative phase relationship between the superposition states \cite{bennett1993teleporting,wang2017generation,chen2020coherent}. Since these Bell states are neither symmetric nor anti-symmetric, their two-photon quantum interference patterns merely provide limited or incomplete scientific and practical significance for high-dimensional quantum information processing. In this work, we challenge this ``well-known'' formulation and show that an alternative complete Bell basis can behave as the symmetric and anti-symmetric quantum states involving multiple spatial modes. Assisted by the two-photon quantum interference that works as a precise Bell state filter, we experimentally prepare the four-dimensional Bell states and verify them by state tomography conditioning on the coincidence counts in two distinct spatial modes. In addition, by varying the relative delay that the paired photons arrive at the quantum interferometry \cite{hikkam2021high}, we are allowed to directly observe the high-dimensional two-photon interference patterns in spatial modes, which manifest themselves as dip or peak within a Gaussian temporal envelope resulting from the bunching and anti-bunching effects. These results confirm that
our scheme has the potential to enable high-dimensional quantum teleportation based on Bell state filters, high-dimensional superdense coding for quantum communication \cite{williams2017superdense,hu2020experimental,luo2019quantum}, as well as quantum metrology that based on nonlocal measurement \cite{mitchell2003diagnosis}, and may also indicate a new direction towards fully harnessing two-photon quantum in more complex high-dimensional quantum information processing \cite{wang2015quantum,erhard2020advances,cozzolino2019high}.

\section{Results}
\subsection{Preparation of four-dimensional Bell states through quantum interference}
Since the existing generalised d-dimensional Bell states of a bipartite system, that defined as $\ket{\phi}_{m,n}=\frac{1}{\sqrt{d}}\sum_{k=0}^{d-1}e^{i2\pi nk/d}\ket{k}_A\ket{m\oplus k}_B$, are obviously neither symmetric nor anti-symmetric, they are not suitable for the direct observation of high-dimensional quantum interference, and as well as those quantum technologies based on quantum interference. Here, we present an alternative complete Bell basis that is encoded in OAM modes $\ket{\ell_1}$, $\ket{\ell_2}$, $\ket{\ell_3}$, and $\ket{\ell_4}$, whose specific formulation can be expressed as
\begin{equation}
\begin{split}
\ket{\psi}_{m,n}^1=\frac{\ket{\ell_1}\ket{\ell_4}+e^{im\pi}\ket{\ell_4}\ket{\ell_1}+e^{in\pi}\ket{\ell_2}\ket{\ell_3}+e^{i(m+n)\pi}\ket{\ell_3}\ket{\ell_2}}{2}\\
\ket{\psi}_{m,n}^2=\frac{\ket{\ell_1}\ket{\ell_3}+e^{im\pi}\ket{\ell_3}\ket{\ell_1}+e^{in\pi}\ket{\ell_2}\ket{\ell_4}+e^{i(m+n)\pi}\ket{\ell_4}\ket{\ell_2}}{2}\\
\ket{\psi}_{m,n}^3=\frac{\ket{\ell_1}\ket{\ell_2}+e^{im\pi}\ket{\ell_2}\ket{\ell_1}+e^{in\pi}\ket{\ell_3}\ket{\ell_4}+e^{i(m+n)\pi}\ket{\ell_4}\ket{\ell_3}}{2}\\
\ket{\psi}_{m,n}^4=\frac{\ket{\ell_1}\ket{\ell_1}+e^{im\pi}\ket{\ell_2}\ket{\ell_2}+e^{in\pi}\ket{\ell_3}\ket{\ell_3}+e^{i(m+n)\pi}\ket{\ell_4}\ket{\ell_4}}{2},
\end{split}
\end{equation}
where $m,n\in\{0,1\}$ such that each group $\ket{\psi}_{m,n}^{1,2,3,4}$ corresponds to four different Bell states, and thus these all sixteen Bell state are mutually orthogonal that are sufficient to constitute a complete Bell basis in two-photon four-dimensional Hilbert space (see appendixes for more details). The most significant aspect of these Bell states is that their symmetric or anti-symmetric properties can be deterministically distinguished, wherein the six quantum states $\ket{\psi}_{1,n}^{1,2,3}$ are anti-symmetric, and the remaining ten quantum states $\ket{\psi}_{0,n}^{1,2,3}$ and $\ket{\psi}_{m,n}^{4}$ are anti-symmetric. Therefore, we can predict that the HOM interference for these sixteen Bell states would behave as the expected and well-known HOM-dips and HOM-peaks.

\begin{figure}[htbp]
\centering
\includegraphics[width=1\linewidth]{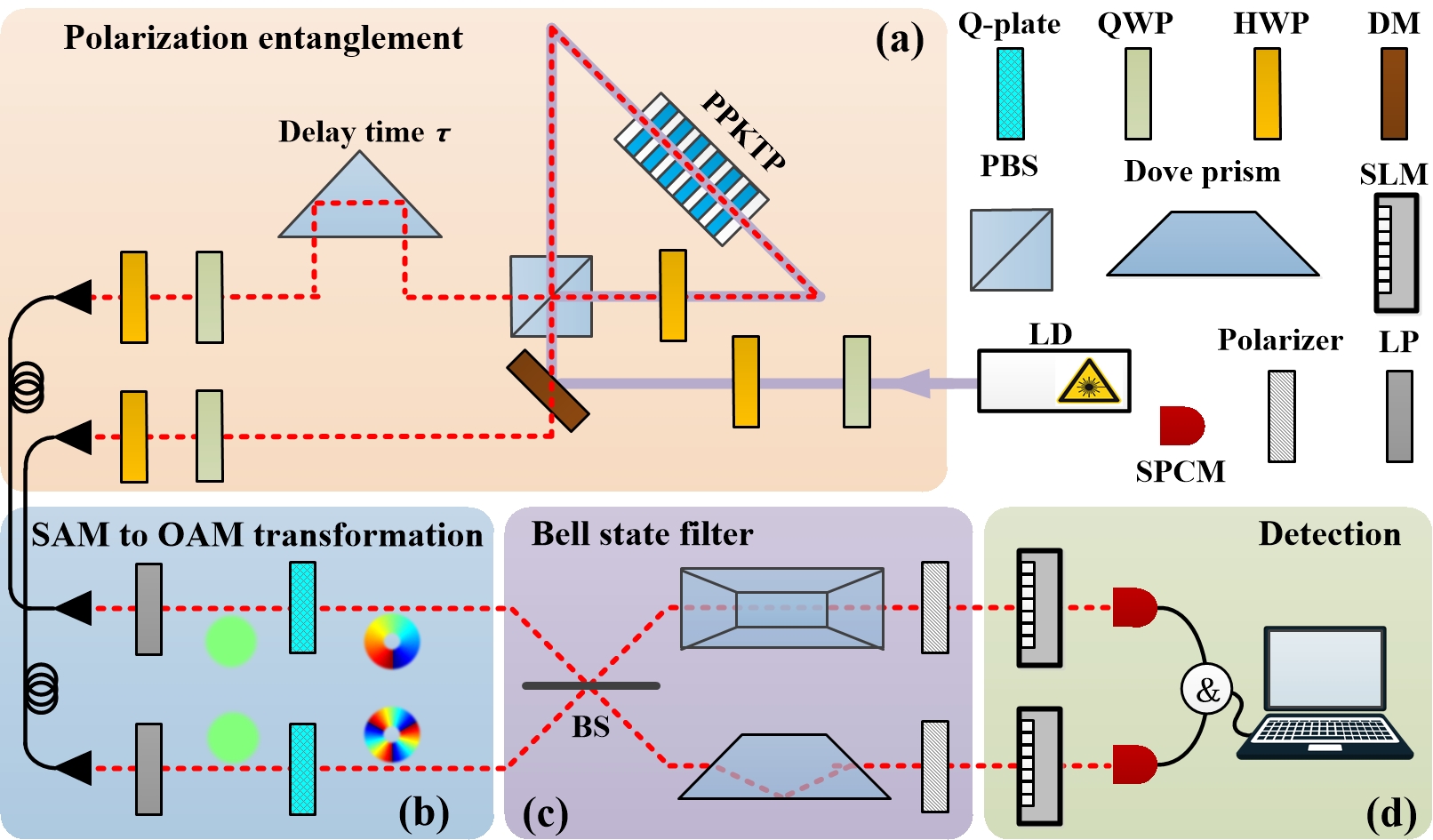}
\vspace{-6mm}
\caption{The experimental setup for harnessing four-dimensional Bell states via quantum interference primarily consists of four modules: (a) polarization entanglement preparation, (b) SAM to OAM transformation based on spin-orbit coupling, (c) a Bell state filter based on HOM interference, and (d) a spatial mode projection detection. Q-plate Vortex Half-Wave Plate, QWP quarter-wave plate, HWP half-wave plate, DM dichroic mirror, SPDM single photon detection module, PPKTP type-II periodically poled potassium titanyl phosphate crystal, PBS polarizing beam splitter, BS beam splitter, SLM  spatial light modulator, LP long pass filter.}
\label{figure_1}
\end{figure}
However, these sixteen Bell states cannot be prepared by neither performing Pauli-X operation because the correlation between OAM modes of paired photons are not cyclic transferred, nor adaptive pump modulation because the relative phase relationship between the superposition states cannot be introduced by adding a common-used Dove prism. To tackle this issue, we exploit polarization-to-OAM transformation and Bell state filter to generate these Bell states in a tunable method. As shown in Fig.\ \ref{figure_1}, a type-II PPKTP nonlinear crystal is placed inside a Sagnac interferometer and pumped with a continuous wave grating-stabilized laser. A half-wave plate in the pump beam is used to set a diagonal polarization state, such that both the clockwise and counterclockwise directions of the interferometer are pumped equally. After combing the photon pairs from both directions on the polarization beam splitter, its resultant polarization entangled state is $\ket{\psi}=(\ket{H}\ket{V}+\ket{V}\ket{H})/\sqrt{2}$, where $H$ and $V$ represent the horizontal and vertical polarization. Since the relative phase and polarization state can be tuned by varying the angle of quarter-wave and half-wave plates, we are able to prepare the complete polarization Bell states. \textcolor{black}{Then these polarization-entangled photons are coupled into single mode fibers to spatially filtering them to a Gaussian mode.} By adding two q-plates with different q values as $q_s=\ell_1/2$ and $q_i=\ell_3/2$ into the signal and idler photons' paths respectively, the polarization entanglement is transformed into $\ket{\psi}\rightarrow(\ket{R,\ell_1}\ket{L,-\ell_3}+\ket{L,-\ell_1}\ket{R,\ell_3})/\sqrt{2}$. Inspired by using the quantum interference as a Bell state filter \cite{zhang2016engineering}, we route these paired photons into a balanced beam splitter from opposite input ports, which constitutes a HOM interferometer. The unitary operation of this beam splitter can be demonstrated by a Hadamard matrix, and transforms the initial structured entanglement into the superposition as $\ket{\psi}\rightarrow\ket{\psi}_A+\ket{\psi}_B$, where $\ket{\psi}_A$ and $\ket{\psi}_B$ correspond to the events that two photons emerge in opposite and identical output ports, respectively. The coincidence detection events identified by two detectors at opposite spatial modes merely record the anti-symmetric Bell states. Thus, it has been widely confirmed that only anti-symmetric two-dimensional Bell states can anti-bunch into opposite output ports of the beam splitter, so-called Bell state filter. As a direct result, the entangled state after this quantum operation can be re-written as
\begin{equation}
\begin{split}
\ket{\psi}_A\rightarrow&(\ket{R,\ell_1}\ket{L,-\ell_3}-\ket{L,-\ell_3}\ket{R,\ell_1}\\
&+\ket{L,-\ell_1}\ket{R,\ell_3}-\ket{R,\ell_3}\ket{L,-\ell_1})/2.
\end{split}
\end{equation}
Followed by two polarizers to erase the distinguishability in polarization degree of freedom, we could obtain the Bell state $\ket{\psi}_{1,0}^{1}$, where $\ket{\ell_4}=\ket{-\ell_3}$ and $\ket{\ell_2}=\ket{-\ell_1}$. Single photons can pass through each polarizer with merely half probability, which would result in a totally 75\% drop in the coincidence detection rate.Since the detection post-selection that merely records the anti-bunched photons separated in distinct spatial modes, here we note that the theoretical prediction of successful probability for generating these four-dimensional Bell states is ultimately limited by 50\%. Additionally, since the bunched photons are still present in the beam path, we also note that they would make deleterious contributions to their further quantum interference, and hinder their use in the quantum superdense coding and quantum teleportation applications. To tackle this issue, auxiliary entanglement in another degree of freedom or auxiliary entangled photons are required to eliminate these bunched photons \cite{wang2015quantum,chen2021temporal}. By engineering the incident polarization entanglement into $(\ket{R}\ket{L}+ e^{i\theta} \ket{L}\ket{R})/\sqrt{2}$ and $(\ket{R}\ket{R}+ e^{i\theta} \ket{L}\ket{L})/\sqrt{2}$ by tuning the half-wave and quarter-wave plates, and modulating the relative phase by rotating two Dove prisms, the OAM Bell states $\ket{\psi}_{m,n}^{1,2}$ can be readily obtained. Analogously, by adding a spiral phase plate to modulate the OAM values of single photons, we can also prepare the OAM Bell states $\ket{\psi}_{m,n}^{3,4}$ (see appendixes for more details). Therefore, through the combination of the tunable controlling of polarization entanglement and the HOM operation for a Bell state filter, we are allowed to generate the complete four-dimensional OAM Bell basis.
\begin{figure}[htbp]
\centering
\includegraphics[width=0.95\linewidth]{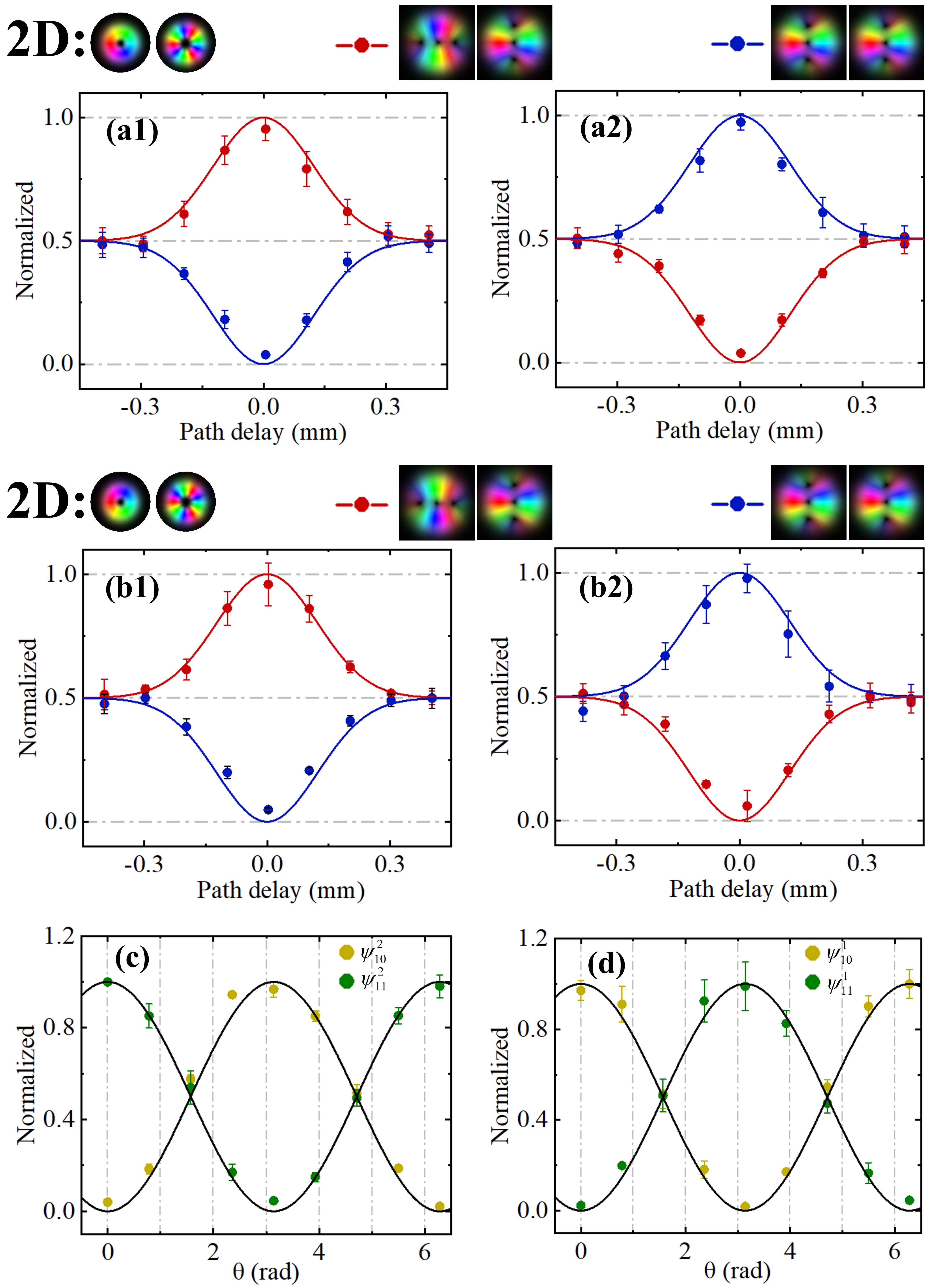}
\vspace{-4mm}
\caption{Experimental observation of HOM interference in two-dimensional subspaces. The circular insets show the input spatial modes of paired photons, and the square insets show their projected measurement bases. (a1-2) HOM interference pattern when the signal photons are projected onto $(\ket{1}+\ket{3})/\sqrt{2}$, and its partner idler photons are projected onto $(\ket{1}-\ket{3})/\sqrt{2}$ (red dots) and $(\ket{1}+\ket{3})/\sqrt{2}$ (blue dots). (b1-2) HOM interference pattern when the signal photons are projected onto $(\ket{-1}+\ket{-3})/\sqrt{2}$, and its partner idler photons are projected onto $(\ket{-1}-\ket{-3})/\sqrt{2}$ (red dots) and $(\ket{-1}+\ket{-3})/\sqrt{2}$ (blue dots). (c-d) The coincidence detection when the paired photons are projected onto the measurement bases $(\ket{1}+\ket{-1})/\sqrt{2}$ and $(\ket{3}+e^{i\theta}\ket{-3})/\sqrt{2}$. The error bars are standard deviations calculated from multiple consecutive measurements and the curves are calculated from the theoretically expected visibilities.}
\label{figure_2}
\end{figure}

\begin{figure*}[htbp]
\centering
\includegraphics[width=1\linewidth]{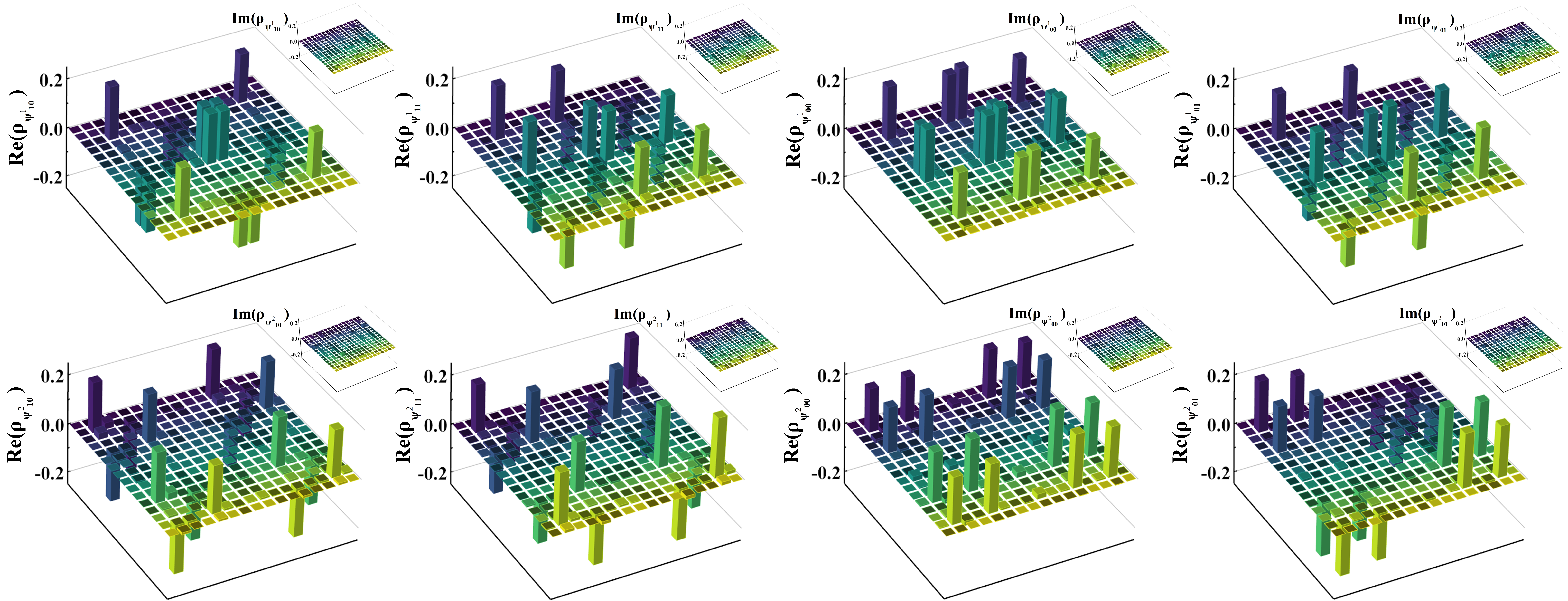}
\vspace{-5mm}
\caption{The density matrices for eight Bell states $\ket{\psi}_{1,0}^{1,2}$, $\ket{\psi}_{1,1}^{1,2}$ $\ket{\psi}_{0,0}^{1,2}$ and $\ket{\psi}_{0,1}^{1,2}$. The labels would read $\ket{-3} \ket{-3}$, $\ket{-3} \ket{-1}$, $\ket{-3} \ket{1}$, $\ket{-3} \ket{3}$,  $\ket{-1} \ket{-3}$, $\ket{-1} \ket{-1}$, $\ket{-1} \ket{1}$, $\ket{-1} \ket{3}$,  $\ket{1} \ket{-3}$, $\ket{1} \ket{-1}$, $\ket{1} \ket{1}$, $\ket{1} \ket{3}$  $\ket{3} \ket{-3}$, $\ket{3} \ket{-1}$, $\ket{3} \ket{1}$ and $\ket{3} \ket{3}$.}
\label{figure_3}
\end{figure*}
\begin{figure}[hbtp]
\centering
\includegraphics[width=0.9\linewidth]{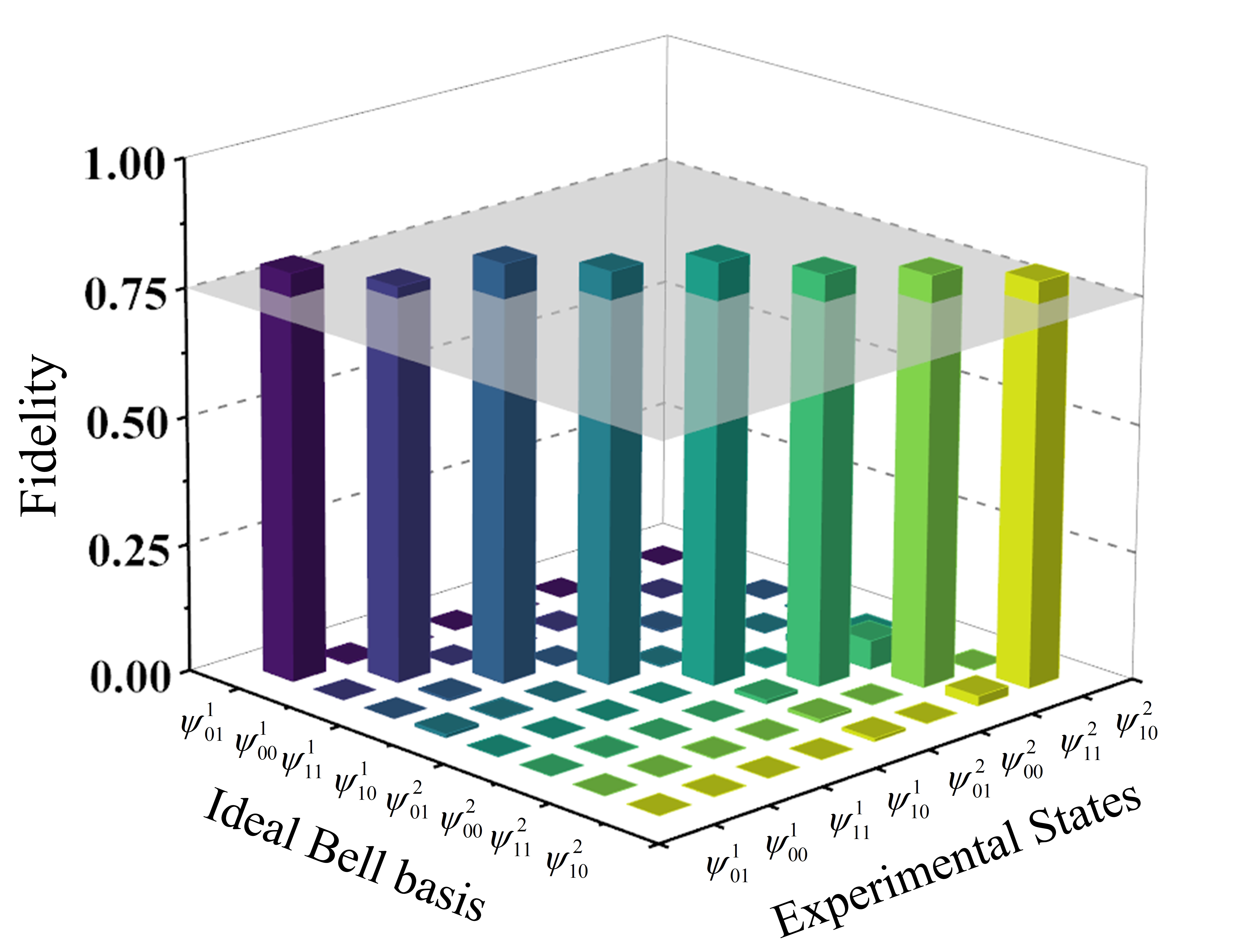}
\vspace{-4mm}
\caption{Fidelity of the experimentally generated state to the ideal Bell basis. All of the generated Bell states in our experiment exceed the theoretical upper bound for the overlap with a three-dimensional entangled state (gray plane).}
\label{figure_4}
\end{figure}
\begin{figure*}[htbp]
\centering
\includegraphics[width=0.85\linewidth]{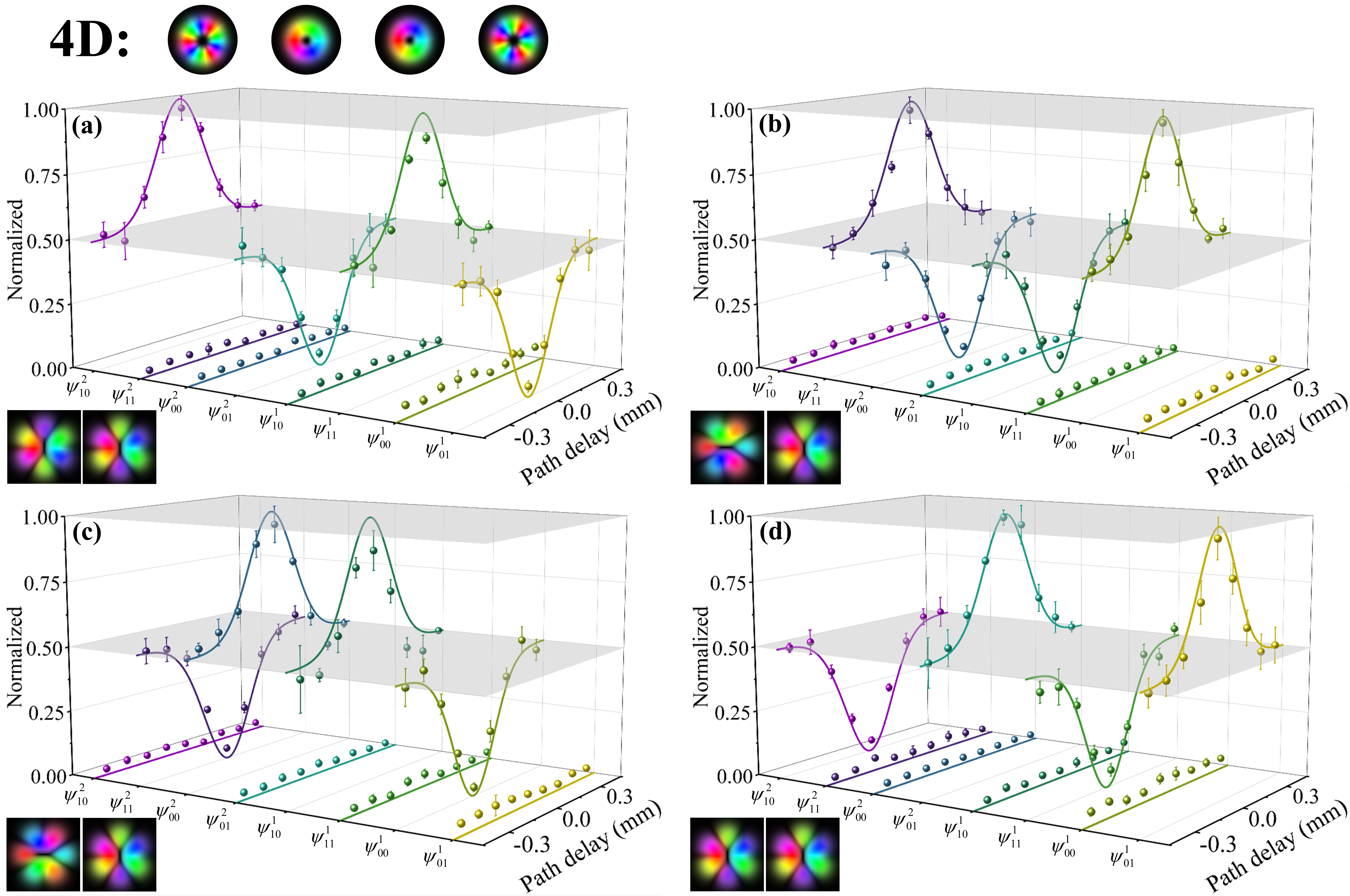}
\vspace{-2mm}
\caption{Two-photon interference in four-dimensional Hilbert space $\ket{\pm 1}\&\ket{ \pm 3}$ in terms of eight experimentally generated Bell states in (a) $\ket{-3}-\ket{-1}-\ket{1}-\ket{3}$ and $\ket{-3}+\ket{-1}+\ket{1}-\ket{3}$ mode pairs, (b) in $\ket{-3}-\ket{-1}+\ket{1}+\ket{3}$ and $\ket{-3}+\ket{-1}+\ket{1}-\ket{3}$ mode pairs, (c) in $\ket{-3}+\ket{-1}-\ket{1}+\ket{3}$ and $\ket{-3}+\ket{-1}+\ket{1}-\ket{3}$ mode pairs, and (d) in $\ket{-3}+\ket{-1}+\ket{1}-\ket{3}$ and $\ket{-3}+\ket{-1}+\ket{1}-\ket{3}$ mode pairs filtered separately. The circular insets show the input spatial modes of paired photons, and the square insets show their projected measurement bases. The error bars are standard deviations calculated from multiple consecutive measurements and the curves are calculated from the theoretically expected visibilities with a Gaussian function.}
\label{figure_5}
\end{figure*}
In our experimental implementation, we use two q-plates with $q_s=1/2$ and $q_i=3/2$ to prepare the four orthogonal spatial modes as $\ket{\ell_1}=\ket{-3}$, $\ket{\ell_2}=\ket{3}$, $\ket{\ell_3}=\ket{-1}$ and $\ket{\ell_4}=\ket{1}$. Since the HOM operation works as a Bell state filter such that merely the detection of anti-symmetric two-dimensional Bell states are recorded as the coincidence events, we can obtain two different forms of sub-entanglements $(\ket{3}\ket{1}-\ket{1}\ket{3})/\sqrt{2}$ and $(\ket{-3}\ket{-1}-\ket{-1}\ket{-3})/\sqrt{2}$ when the input two-photon polarization states are set as $\ket{R}\ket{R}$ and $\ket{L}\ket{L}$ respectively. To verify the performance of this Bell state filter, quantum interference patterns are directly observed in their two-dimensional subspaces by tuning the relative time delay between paired photons. Here, the projective measurements are performed by using the near-infrared spatial mode modulators (SLM) and single-mode fibers, where the holographic pattern determines the measurement basis. The key to this method lies in employing the SLM to transform the complex spatial modes for efficient fiber coupling, and followed by a single-mode fiber to work as an OAM filter such that merely the expected spatial modes can be routed to the single-photon detector. As shown in Fig. \ref{figure_2}\textcolor{blue}{(a1-b1)}, the peaks are observed if the measurement bases are $(\ket{\pm1}+\ket{\pm3})/\sqrt{2}$ and $(\ket{\pm1}-\ket{\pm3})/\sqrt{2}$, and conversely the dips would be observed if the measurement bases are $(\ket{\pm1}+\ket{\pm3})/\sqrt{2}$ and $(\ket{\pm1}+\ket{\pm3})/\sqrt{2}$. By adding two dove prisms with a relative angle as $45^{\circ}$ in the signal and idler photons' paths to modulate the relative phase between different spatial modes, the symmetric two-dimensional sub-entanglements $(\ket{3}\ket{1}+\ket{1}\ket{3})/\sqrt{2}$ and $(\ket{-3}\ket{-1}+\ket{-1}\ket{-3})/\sqrt{2}$ are obtained. In this case, their interference patterns are shown in Fig. \ref{figure_2}\textcolor{blue}{(a2-b2)}, which are completely reversed with respect to Fig. \ref{figure_2}\textcolor{blue}{(a1-b1)}. According to the measurement results  presented in Fig. \textcolor{blue}{2(a-b)}, the experimentally measured interference visibility in the two-dimensional Hilbert space is $96.0\pm0.4\%$. Analogously, when the input two-photon polarization states are set as $\ket{R}\ket{L}$ and $\ket{L}\ket{R}$, two-dimensional sub-entanglements $(\ket{3}\ket{-1}\pm\ket{-1}\ket{3})/\sqrt{2}$ and $(\ket{-3}\ket{1}\pm\ket{1}\ket{-3})/\sqrt{2}$ are obtained, and their quantum interference patterns also manifest themselves as peaks and dips (see appendixes for more details). Next let us verify the performance of polarization entanglement in our generation process of four-dimensional Bell states. If the initial polarization entanglement is $(\ket{R}\ket{R}\pm\ket{L}\ket{L})/\sqrt{2}$, we can observe the cosine and sinusoidal curves as a function of phase parameter $\theta$ that defined in the measurement bases as $(\ket{1}+\ket{-1})/\sqrt{2}$ and $(\ket{3}+e^{i\theta}\ket{-3})/\sqrt{2}$ as shown in Fig. \ref{figure_2}\textcolor{blue}{(c)}. Analogously, when the initial polarization entanglement is $(\ket{R}\ket{L}\pm \ket{L}\ket{R})/\sqrt{2}$, the corresponding sinusoidal and cosine curves are shown in Fig. \ref{figure_2}\textcolor{blue}{(d)}. Backed by the experimental verification of the tunable controlling of polarization entanglement and the HOM operation for a Bell state filter, we can exploit their combination to prepare the complete four-dimensional symmetric and anti-symmetric Bell states.

Quantum state tomography enables the construction of the density matrix $\rho_{exp}$ \cite{agnew2011tomography} by minimizing the square root of the infidelity, which can be effectively solved by a type of convex optimization problem known as semidefinite programming (SDP). In our experiment, we demonstrate that the preparation of eight four-dimensional Bell states can be expressed as
\begin{equation}
\begin{split}
\ket{\psi}_{1,0}^1=(\ket{-3}\ket{1}-\ket{1}\ket{-3}+\ket{3}\ket{-1}-\ket{-1}\ket{3})/2\\
\ket{\psi}_{1,1}^1=(\ket{-3}\ket{1}-\ket{1}\ket{-3}-\ket{3}\ket{-1}+\ket{-1}\ket{3})/2\\
\ket{\psi}_{0,0}^1=(\ket{-3}\ket{1}+\ket{1}\ket{-3}+\ket{3}\ket{-1}+\ket{-1}\ket{3})/2\\
\ket{\psi}_{0,1}^1=(\ket{-3}\ket{1}+\ket{1}\ket{-3}-\ket{3}\ket{-1}-\ket{-1}\ket{3})/2\\
\\
\ket{\psi}_{1,0}^2=(\ket{-3}\ket{-1}-\ket{-1}\ket{-3}+\ket{1}\ket{3}-\ket{3}\ket{1})/2\\
\ket{\psi}_{1,1}^2=(\ket{-3}\ket{-1}-\ket{-1}\ket{-3}-\ket{1}\ket{3}+\ket{3}\ket{1})/2\\
\ket{\psi}_{0,0}^2=(\ket{-3}\ket{-1}+\ket{-1}\ket{-3}+\ket{1}\ket{3}+\ket{3}\ket{1})/2\\
\ket{\psi}_{0,1}^2=(\ket{-3}\ket{-1}+\ket{-1}\ket{-3}-\ket{1}\ket{3}-\ket{3}\ket{1})/2.
\end{split}
\end{equation}
Figure \ref{figure_3} displays all the reconstructed density matrices of the four-dimensional Bell states generated in the experiment. In order to assess the quality of the generated Bell states, we evaluate their fidelity to the theoretically expected Bell basis and certify their entanglement dimensionality by a concise yet faithful witness. Fidelity, denoted as $F=\left[\operatorname{Tr}\left(\sqrt{\sqrt{\rho_{T}} \rho_{exp} \sqrt{\rho_{T}}}\right)\right]^2$ where $\rho_{T}$ is the target density matrix, measures the proximity between the reconstructed state and a target state. The average fidelity to the ideal Bell basis for the eight experimentally generated states is $0.80 \pm 0.01$. Fig. \ref{figure_4} displays the fidelity witnesses measured for these Bell states, all of which surpass the upper bound $\bar{F}_{\max }=75 \%$. As a result, we can adequately verify that every Bell state produced in our experiment is entangled with at least four dimensions.  \textcolor{black}{The decrease of the experimentally measured fidelity can be mainly attributed to the imperfect performance of quantum interference, and the slight deviations are caused by the intermodal crosstalk between different OAM modes, imperfect misalignment, and nonideal optical components.} As the commonly-used quantum states $\ket{\psi}_{m,n}^4$ has been investigated \cite{hikkam2021high}, we merely present the experimental results for eight intriguing four-dimensional Bell states in this work for the sake of experimental simplicity. But we note again that the complete high-dimensional Bell states can be prepared by our method with the current technologies (see appendixes for more details).

While our work merely report the experimental observation of two-photon interference in four-dimensional Hilbert space, the whole scheme can be extended to higher-dimensional entanglement by using, for example, path-to-OAM transformation. Additionally, since the entanglement transformation is performed in the form of monogamy, it is allowed to modulate each OAM modes independently, such that quantum entanglement with large OAM quanta and with perfect vortex can be readily prepared \cite{fickler2016quantum,fickler2012quantum,ostrovsky2013generation}. As the bunching probability of HOM interference is directly related to both of the photons' level of indistinguishability and its degree of purity, this technique has the potential to achieve a probabilistic distillation of a Bell singlet state even when the incident photons are subjected to the action of local noisy environments \cite{piccolini2023robust,ollivier2021hong}.

\subsection{High-dimensional two-photon interference effects in spatial modes}
Benefiting from the inherent advantages of spatial modes like controllable manipulation and flexible filtering, we investigate the high-dimensional interference during the generation and detection of superposition states, which is still a challenging task to be implemented in other degrees of freedom. Now let we use a concise yet efficient scan of temporal delay to directly observe two-photon interference effects in four-dimensional Hilbert space $\ket{\pm 1}\&\ket{ \pm 3}$ in terms of eight experimentally generated Bell states. In the experimental generation process, we directly tune the relative time delay between different paths of the interferometer. In the output paths, the measurement module for detecting OAM modes uses the SLM to make an expected change to the incident OAM modes, and followed by a single mode fiber such that only those photons with $\ell=0$ can be detected. As a direct result, we can observe the high-dimensional two-photon interference patterns under different projection measurement bases as a function of the relative time delay.

As shown in Fig. \ref{figure_5}\textcolor{blue}{(a)}, if the two measurement bases are set as $\ket{-3}-\ket{-1}-\ket{1}-\ket{3}$ and $\ket{-3}+\ket{-1}+\ket{1}-\ket{3}$, we can observe the peaks for Bell states $\ket{\psi}_{1,1}^1$ and $\ket{\psi}_{1,0}^2$ as a direct manifestation of anti-bunching effect. Conversely, the dips are observed for Bell states $\ket{\psi}_{0,1}^1$ and $\ket{\psi}_{0,1}^2$ as a direct manifestation of bunching effect. On the other hand, no twofold coincidence events are detected for Bell states $\ket{\psi}_{0,0}^1$, $\ket{\psi}_{1,0}^1$, $\ket{\psi}_{0,0}^2$ and $\ket{\psi}_{1,1}^2$ since they and the measurement bases are completely orthogonal. Analogously, various two-photon interference patterns are presented in in Fig. \ref{figure_5}\textcolor{blue}{(b-d)} for different measurement bases. To quantify the performance of two-photon interference, we utilize the visibility defined as $V=|1-{C}_{\tau=0}/{C}_{\tau=\infty}|$ for the observed dips or peaks. For these eight Bell states generated in our experiment, the average visibility of the two-photon interference in the four-dimensional Hilbert space is as high as $90.31\pm5.52\%$.

\section{Conclusions}
We have demonstrated a generalised formulation of high-dimensional symmetric and anti-symmetric Bell states, and experimentally prepared four-dimensional OAM Bell states by using quantum interference.
Our work uses the impact of state symmetry on the HOM effect to filter the specific anti-symmetric Bell states in multiple two-dimensional Hilbert spaces without any post-selection other than conditioning on coincidences. This makes our approach can be readily extended to higher-dimensional entanglement, and other forms of entanglement like temporal modes. \textcolor{black}{For example, integrated photonic chips have recently emerged as a leading platform for the harnessing entangled photons, wherein the high-dimensional path entanglement can be readily tunable. Building on the principle of our method, these path-entangled photons can be used to prepare the complete high-dimensional OAM Bell states based on the combination of path-to-OAM transformation and HOM interference that behaving as Bell state filter.} While our experimental realization is based on bulk optics, the overall approach could also be extended to integrated quantum optics, which has the potential to be applied in more complex quantum computation tasks.

Although the two-photon quantum interference in high-dimensional spatial modes has been reported \cite{hikkam2021high,zhang2016engineering,zhang2017simultaneous}, none to date have demonstrated high-dimensional interference for the complete Bell basis. However, the complete Bell basis and its quantum interference are essential prerequisite for many quantum information processing applications \cite{cozzolino2019high,erhard2020advances}. To overcome this obstacle, our work provides an alternative approach to harness high-dimensional symmetric and anti-symmetric Bell state through quantum interference.
\textcolor{black}{While the complete high-dimensional Bell states can be generated by various methods, like using single-photon quantum gate or tailoring spatial mode of the pump beam, the implementation complexity of cyclic transformation exponentially increases with increasing entanglement dimensionality, and the efficiency of SPDC for creating entangled photons significantly decreases with respect to higher order spatial mode of the pump beam. As a direct result, the experimental creation of an arbitrary dimensional Bell basis puts a brake on the development of high-dimensional quantum information processing. Compared with these existing methods, our method aims at preparing the complete high-dimensional Bell states that are all symmetric or anti-symmetric quantum states, which is an essential property for high-dimensional information processing protocols based on quantum interference. Additionally, since our method builds on the entanglement transformation and HOM interference, it eludes the requirements for complex single-photon quantum gate and inefficient coherent generation by adaptive pump modulation. However, the entanglement transformation between different degrees of freedom maybe a technical challenge for preparing higher-dimensional Bell states. As have stated, the path entanglement in integrated photonic chips has the potential to tackle this issue of entanglement transformation that is necessary in our method.}

Our results indicates that the interference patterns for six anti-symmetric Bell states or ten symmetric Bell states are identical respectively. With the assistance of quantum non-demolition measurement for the elimination of the deleterious bunched photons \cite{wang2015quantum,chen2021temporal}, we note that the complete high-dimensional Bell state measurement may be realized by the combination of HOM interference and auxiliary entanglement or photons \cite{luo2019quantum,hu2020experimental,wang2022four}. Additionally, our method may also provide entanglement sources for some potential applications that have no requirement for the recombination of paired photons, such as those quantum information protocols based on single-photon Bell-state measurement \cite{barreiro2008beating,li2020quantum,liang2015simple}.
We believe that fully harnessing quantum interference can provide valuable tools for engineering complex forms of high-dimensional multiphoton entanglement, as required for the next generation of large-scale quantum information processing protocols.

\begin{acknowledgments}
This work is supported by the National Natural Science Foundation of China (NSFC) (12034016, 12004318, 61975169), the Fundamental Research Funds for the Central Universities at Xiamen University (20720190057, 20720210096), the Natural Science Foundation of Fujian Province of China (2020J05004), the Natural Science Foundation of Fujian Province of China for Distinguished Young Scientists (2015J06002), and the program for New Century Excellent Talents in University of China (NCET-13-0495).
\end{acknowledgments}

\section*{APPENDIX A: Experimental details}
We employ a $5mm$ nonlinear PPKTP crystal with a $10.025um$ grating period to achieve type-II quasi-phase-matched collinear parametric down-conversion, resulting in the generation of paired photons with degenerate wavelength at $810 nm$. The polarization modes of the paired photons are entangled through bidirectional down-conversion by placing the nonlinear crystal in a Sagnac loop. As a direct result of this Sagnac interferometer, the produced photons can be guided with certainty into two separate spatial modes. By utilizing polarization and phase modulation for the incident pump light, it is allowed to achieve all four maximally polarization-entangled Bell states. Following the Sagnac-type entanglement source, paired photons are directed into separated single-mode fibers, where their spatial modes undergo appropriate filtering to eliminate complex structured modes and obtain a simple Gaussian mode. The polarization-entangled states can be converted into OAM states by introducing a sequence of components, including two quarter-wave plates and a Q-plate with topological charges of $q=m/2$ in the signal and idler paths. Subsequently, the entangled structured photons are directed to a balanced beam splitter from two opposite input ports, which implements the HOM operation for a Bell state filter. Followed by modulating the relative phase through rotating two Dove prisms and erasing the distinguishability in polarization degree of freedom through two polarizers, the OAM Bell states $\ket{\psi}_{m,n}^{1,2}$ can be readily obtained. Afterwards, two spatial light modulators and single mode fibers are used to extract a specific spatial transverse mode. Finally, the silicon avalanche photon diodes capture the paired photons, and the rapid electronic AND gate identifies two-fold coincidence events when both photons reach the separated detectors within a coincidence window of approximately $1ns$.

Since this quantum interference should behave identically for different photonic modes, the relative time delay between paired photons must be fixed even when the photons with orthogonal OAM or polarization modes may carry some deleterious dispersion caused by the optical components or noisy environment. For instance, the polarization instruments may introduce a time delay between horizontal and vertical polarized modes. Additionally, the photons with different OAM quanta would suffer from tiny difference in their propagation speed, and thus if the distance that the photon travels is long enough, it may introduce a temporal shift in the HOM interferometry. In our experiment, we measure that the path delay between sub-entanglements $(\ket{-1}\ket{-3}-\ket{-3}\ket{-1})/\sqrt{2}$ and $(\ket{1}\ket{3}-\ket{3}\ket{1})/\sqrt{2}$ is approximately $0.07 mm$, and thus two quartz crystals with thickness of $8 mm$ are added to compensate this time delay such that the quantum interference visibilities are simultaneously high for both two-dimensional subspaces.

Since our scheme utilizes quantum interference as a Bell filter that can simultaneously distil singlet states in multiple two-dimensional Hilbert space, the dimensions of the symmetric and anti-symmetric Bell states are preferred to be even number. For example, let us consider a three-dimensional path-entangled state as $\ket{\psi}_p=(\ket{a}\ket{a}+\ket{b}\ket{b}+\ket{c}\ket{c})/\sqrt{3}$. By placing the q-plates or spiral phase plates in each spatial modes to achieve the path-to-OAM transformation as $\ket{\psi}_p\rightarrow(\ket{a,\ell_1}\ket{a,\ell_2}+\ket{b,\ell_3}\ket{b,\ell_4}+\ket{c,\ell_5}\ket{c,\ell_6})/\sqrt{3}$. Then two beam combiners in signal and idler paths are used to erase the distinguishability in spatial modes, and route these paired photons into a HOM interferometer. The resultant Bell state becomes
\begin{equation}
\begin{split}
\ket{\psi}\rightarrow&(\ket{\ell_1}\ket{\ell_2}-\ket{\ell_2}\ket{\ell_1}+\ket{\ell_3}\ket{\ell_4}\\
&-\ket{\ell_4}\ket{\ell_3}+\ket{\ell_5}\ket{\ell_6}-\ket{\ell_6}\ket{\ell_5})/\sqrt{6},
\end{split}\tag{A1}
\end{equation}
which is an anti-symmetric Bell state in six-dimensional Hilbert space. Analogously, by engineering the OAM correlation between paired photons through q-plates and spiral phase plates, and modulating their relative phases through dove prisms and quarter waveplates that can introduce OAM and polarization mode-dependent phase respectively, the complete six-dimensional Bell basis, all of which are either symmetric or anti-symmetric, can be obtained with merely linear optics.

\section*{APPENDIX B: Generation of the complete four-dimensional Bell basis}
\begin{table*}[htbp]
\caption{Transformations for the complete four-dimensional symmetric and anti-symmetric OAM Bell states.}
\label{tab1}
\begin{tabular}{p{4.5cm}<{\centering}p{1.2cm}<{\centering}p{1.2cm}<{\centering}p{1.2cm}<{\centering}p{1.5cm}<{\centering}p{6.5cm}<{\centering}}
\hline
\hline
Initial polarization state $\otimes\ket{a}\ket{b}$ & Q-plate & SPP & Dove & Flip OAM & Final Bell state $\otimes\ket{c}\ket{d}$\\
\hline
$\ket{R}\ket{L}+\ket{L}\ket{R})/\sqrt{2}$ &$m_i=1$ $m_s=3$& No &$\alpha=0$& No & $\ket{\psi}_{1,0}^1=(\ket{-3}\ket{1}-\ket{1}\ket{-3}+\ket{3}\ket{-1}-\ket{-1}\ket{3})/2$ \\
\hline
$\ket{R}\ket{L}-\ket{L}\ket{R})/\sqrt{2}$ &$m_i=1$ $m_s=3$& No &$\alpha=0$& No & $\ket{\psi}_{1,1}^1=(\ket{-3}\ket{1}-\ket{1}\ket{-3}-\ket{3}\ket{-1}+\ket{-1}\ket{3})/2$ \\
\hline
$\ket{R}\ket{L}+i\ket{L}\ket{R})/\sqrt{2}$ &$m_i=1$ $m_s=3$& No &$\alpha=\pi/8$& No & $\ket{\psi}_{0,0}^1=(\ket{-3}\ket{1}+\ket{1}\ket{-3}+\ket{3}\ket{-1}+\ket{-1}\ket{3})/2$ \\
\hline
$\ket{R}\ket{L}-i\ket{L}\ket{R})/\sqrt{2}$ &$m_i=1$ $m_s=3$& No &$\alpha=\pi/8$& No & $\ket{\psi}_{0,1}^1=(\ket{-3}\ket{1}+\ket{1}\ket{-3}-\ket{3}\ket{-1}-\ket{-1}\ket{3})/2$ \\
\hline

$\ket{R}\ket{R}-\ket{L}\ket{L})/\sqrt{2}$ &$m_i=1$ $m_s=3$& No &$\alpha=0$& No & $\ket{\psi}_{1,0}^2=(\ket{-3}\ket{-1}-\ket{-1}\ket{-3}+\ket{1}\ket{3}-\ket{3}\ket{1})/2$ \\
\hline
$\ket{R}\ket{R}+\ket{L}\ket{L})/\sqrt{2}$ &$m_i=1$ $m_s=3$& No &$\alpha=0$& No & $\ket{\psi}_{1,1}^2=(\ket{-3}\ket{-1}-\ket{-1}\ket{-3}-\ket{1}\ket{3}+\ket{3}\ket{1})/2$ \\
\hline
$\ket{R}\ket{R}-\ket{L}\ket{L})/\sqrt{2}$ &$m_i=1$ $m_s=3$& No &$\alpha=\pi/4$& No & $\ket{\psi}_{0,0}^2=(\ket{-3}\ket{-1}+\ket{-1}\ket{-3}+\ket{1}\ket{3}+\ket{3}\ket{1})/2$ \\
\hline
$\ket{R}\ket{R}+\ket{L}\ket{L})/\sqrt{2}$ &$m_i=1$ $m_s=3$& No &$\alpha=\pi/4$& No & $\ket{\psi}_{0,1}^2=(\ket{-3}\ket{-1}+\ket{-1}\ket{-3}-\ket{1}\ket{3}-\ket{3}\ket{1})/2$ \\
\hline

$\ket{L}\ket{R}+\ket{R}\ket{L})/\sqrt{2}$ &$m_i=1$ $m_s=1$& $l_i=2$ $l_s=-2$ &$\alpha=0$& No & $\ket{\psi}_{1,0}^3=(\ket{-3}\ket{3}-\ket{3}\ket{-3}+\ket{-1}\ket{1}-\ket{1}\ket{-1})/2$ \\
\hline
$\ket{L}\ket{R}-\ket{R}\ket{L})/\sqrt{2}$ &$m_i=1$ $m_s=1$& $l_i=2$ $l_s=-2$ &$\alpha=0$& No & $\ket{\psi}_{1,1}^3=(\ket{-3}\ket{3}-\ket{3}\ket{-3}-\ket{-1}\ket{1}+\ket{1}\ket{-1})/2$ \\
\hline
$\ket{L}\ket{R}-\ket{R}\ket{L})/\sqrt{2}$ &$m_i=1$ $m_s=1$& $l_i=2$ $l_s=-2$ &$\alpha=\pi/4$& No & $\ket{\psi}_{0,0}^3=(\ket{-3}\ket{3}+\ket{3}\ket{-3}+\ket{-1}\ket{1}+\ket{1}\ket{-1})/2$ \\
\hline
$\ket{L}\ket{R}+\ket{R}\ket{L})/\sqrt{2}$ &$m_i=1$ $m_s=1$& $l_i=2$ $l_s=-2$ &$\alpha=\pi/4$& No & $\ket{\psi}_{0,1}^3=(\ket{-3}\ket{3}+\ket{3}\ket{-3}-\ket{-1}\ket{1}-\ket{1}\ket{-1})/2$ \\
\hline

$\ket{L}\ket{R}+\ket{R}\ket{L})/\sqrt{2}$ &$m_i=1$ $m_s=1$& $l_i=2$ $l_s=-2$ &$\alpha=0$& $ Yes $ & $\ket{\psi}_{1,0}^4=(\ket{-3}\ket{-3}-\ket{3}\ket{3}+\ket{-1}\ket{-1}-\ket{1}\ket{1})/2$ \\
\hline
$\ket{L}\ket{R}-\ket{R}\ket{L})/\sqrt{2}$ &$m_i=1$ $m_s=1$& $l_i=2$ $l_s=-2$ &$\alpha=0$& $ Yes $ & $\ket{\psi}_{1,1}^4=(\ket{-3}\ket{-3}-\ket{3}\ket{3}-\ket{-1}\ket{-1}+\ket{1}\ket{1})/2$ \\
\hline
$\ket{L}\ket{R}-\ket{R}\ket{L})/\sqrt{2}$ &$m_i=1$ $m_s=1$& $l_i=2$ $l_s=-2$ &$\alpha=\pi/4$& $ Yes $ & $\ket{\psi}_{0,0}^4=(\ket{-3}\ket{-3}+\ket{3}\ket{3}+\ket{-1}\ket{-1}+\ket{1}\ket{1})/2$ \\
\hline
$\ket{L}\ket{R}+\ket{R}\ket{L})/\sqrt{2}$ &$m_i=1$ $m_s=1$& $l_i=2$ $l_s=-2$ &$\alpha=\pi/4$& $ Yes $ & $\ket{\psi}_{0,1}^4=(\ket{-3}\ket{-3}+\ket{3}\ket{3}-\ket{-1}\ket{-1}-\ket{1}\ket{1})/2$ \\
\hline
\hline
\end{tabular}
\end{table*}
In our experimental scheme, the incident polarization entanglement states is prepared as $\ket{\psi}=(\ket{R_0}\ket{R_0}-\ket{L_0}\ket{L_0})/\sqrt{2}\otimes\ket{a}\ket{b}$, where the spatial mode $\ell=0$ is filter by using two single mode fibers. By placing a series of half and quarter waveplates and vortex plates with $q_s=1/2$ and $q_i=3/2$ in signal and idler paths, respectively, the entangled state is transformed to a hybridentangled state as
\begin{equation}
\ket{\psi}\rightarrow(\ket{R_{1}}\ket{R_{3}}-\ket{L_{-1}}\ket{L_{-3}})/\sqrt{2}\otimes\ket{a}\ket{b}.\tag{B1}
\end{equation}
After the operation of a balanced beam splitter that defined as a Hadamard transformation, the state becomes
\begin{equation}
\begin{split}
\ket{\psi}\rightarrow&(\ket{R_{1}^d}+i\ket{R_{1}^c})\otimes(\ket{R_{3}^c}+i\ket{R_{3}^d})\\
&-(\ket{L_{-1}^d}+i\ket{L_{-1}^c})\otimes(\ket{L_{-3}^c}+i\ket{L_{-3}^d})\\
=&(\ket{R_{1}^d}\ket{R_{3}^c}+i\ket{R_{1}^c}\ket{R_{3}^c}+i\ket{R_{1}^d}\ket{R_{3}^d}-\ket{R_{1}^c}\ket{R_{3}^d})\\
&-(\ket{L_{-1}^d}\ket{L_{-3}^c}+i\ket{L_{-1}^c}\ket{L_{-3}^c}+i\ket{L_{-1}^d}\ket{L_{-3}^d}-\ket{L_{-1}^c}\ket{L_{-3}^d}).
\end{split}\tag{B2}
\end{equation}
Since two single-photon detectors at opposite spatial modes merely recodes the coincident events at the distinct output ports of the beam splitter, the resultant anti-bunched quantum state is projected to
\begin{equation}
\ket{\psi}_A\rightarrow\ket{R_{1}^d}\ket{R_{3}^c}-\ket{R_{1}^c}\ket{R_{3}^d}-\ket{L_{-1}^d}\ket{L_{-3}^c}+\ket{L_{-1}^c}\ket{L_{-3}^d}.\tag{B3}
\end{equation}
By placing two polarizers oriented at diagonal direction to eraser the distinguishable information in polarization degree of freedom, we obtain a four-dimensional Bell basis as
\begin{equation}
\begin{split}
\ket{\psi}_{1,0}^2\rightarrow(\ket{-3}\ket{-1}-\ket{-1}\ket{-3}+\ket{1}\ket{3}-\ket{3}\ket{1})/2\otimes\ket{c}\ket{d}.
\end{split}\tag{B4}
\end{equation}

If the incident polarization entanglement states is prepared as $\ket{\psi}=(\ket{R_0}\ket{R_0}+\ket{L_0}\ket{L_0})/\sqrt{2}\otimes\ket{a}\ket{b}$. By repeating the analogous process, the state after the operation of a balanced beam splitter becomes
\begin{equation}
\begin{split}
\ket{\psi}\rightarrow&(\ket{R_1^d}+i\ket{R_1^c})\otimes(\ket{R_3^c}+i\ket{R_3^d})\\
&+(\ket{L_{-1}^d}+i\ket{L_{-1}^c})\otimes(\ket{L_{-3}^c}+i\ket{L_{-3}^d})\\
=&(\ket{R_1^d}\ket{R_3^c}+i\ket{R_1^c}\ket{R_3^c}+i\ket{R_1^d}\ket{R_3^d}-\ket{R_1^c}\ket{R_3^d})\\
&+(\ket{L_{-1}^d}\ket{L_{-3}^c}+i\ket{L_{-1}^c}\ket{L_{-3}^c}+i\ket{L_{-1}^d}\ket{L_{-3}^d}-\ket{L_{-1}^c}\ket{L_{-3}^d}).
\end{split}\tag{B5}
\end{equation}
Due to the spatial filter by recording the coincidence events at distinct detectors and the polarization filter by placing two polarizers that are oriented at diagonal direction, we obtain a four-dimensional Bell basis as
\begin{equation}
\ket{\psi}_{1,1}^2\rightarrow(\ket{-3}\ket{-1}-\ket{-1}\ket{-3}-\ket{1}\ket{3}+\ket{3}\ket{1})/2\otimes\ket{c}\ket{d}.\tag{B6}
\end{equation}

Next, let us adding two Dove prisms after the HOM filter to implement phase modulation. Two dove prisms with a relative rotation angle $\alpha$ would introduce a relative phase as $\ket{\ell}\rightarrow exp(i2\ell\alpha)\ket{\ell}$, and their image abberation can also be compensated. For example, the effect of this operation on the state $\ket{\psi}_{1,0}^{2}$ and $\ket{\psi}_{1,1}^{2}$ can be expressed as
\begin{equation}
\begin{split}
\ket{\psi}_{1,0}^2\rightarrow(e^{-i2\alpha}\ket{-3}\ket{-1}-e^{-i6\alpha}\ket{-1}\ket{-3}+e^{i6\alpha}\ket{1}\ket{3}-e^{i2\alpha}\ket{3}\ket{1})/2\\
\ket{\psi}_{1,1}^2\rightarrow(e^{-i2\alpha}\ket{-3}\ket{-1}-e^{-i6\alpha}\ket{-1}\ket{-3}-e^{i6\alpha}\ket{1}\ket{3}+e^{i2\alpha}\ket{3}\ket{1})/2\\
\end{split}\tag{B7}
\end{equation}
By setting the relative angle of two dove prisms at angle $\alpha=\pi/4$, the above Bell states are readily transformed to $\ket{\psi}_{0,0}^2$ and $\ket{\psi}_{0,1}^2$.

If the incident polarization states are $\ket{\psi}=(\ket{R_0}\ket{L_0}+ e^{i\theta} \ket{L_0}\ket{R_0})/\sqrt{2}\otimes\ket{a}\ket{b}$, the Bell state $\ket{\psi}_{m,n}^1$ are readily obtained by following the analogous operations as above.

The incident polarization entanglement states is prepared as $\ket{\psi}=(\ket{L_0}\ket{R_0}+\ket{R_0}\ket{L_0})/\sqrt{2}\otimes\ket{a}\ket{b}$. By adding a Q-plate and a spiral phase plate (SPP) with $(q_s=1/2, SPP=2)$ and $(q_i=1/2,SPP=-2)$ in the signal and idler paths, respectively. As a direct result, the entangled state is transformed to a hybridentangled state as
\begin{equation}
\ket{\psi}\rightarrow(\ket{L_1}\ket{R_{-1}}+\ket{R_{3}}\ket{L_{-3}})/\sqrt{2}\otimes\ket{a}\ket{b}.\tag{B8}
\end{equation}
After the operation of a balanced beam splitter, the state becomes
\begin{equation}
\begin{split}
\ket{\psi}\rightarrow&(\ket{L_1^d}+i\ket{L_1^c})\otimes(\ket{R_{-1}^c}+i\ket{R_{-1}^d})\\
&+(\ket{R_{3}^d}+i\ket{R_{3}^c})\otimes(\ket{L_{-3}^c}+i\ket{L_{-3}^d})\\
=&(\ket{L_1^d}\ket{R_{-1}^c}+i\ket{L_1^c}\ket{R_{-1}^c}+i\ket{L_1^d}\ket{R_{-1}^d}-\ket{L_1^c}\ket{R_{-1}^d})\\
&+(\ket{R_{3}^d}\ket{L_{-3}^c}+i\ket{R_{3}^c}\ket{L_{-3}^c}+i\ket{R_{3}^d}\ket{L_{-3}^d}-\ket{R_{3}^c}\ket{L_{-3}^d}).
\end{split}\tag{B9}
\end{equation}
Since two single-photon detectors at distinct spatial modes merely recodes the coincident at the distinct output ports of the beam splitter, the resultant quantum state is projected to
\begin{equation}
\ket{\psi}\rightarrow\ket{L_1^d}\ket{R_{-1}^c}-\ket{L_1^c}\ket{R_{-1}^d}+\ket{R_{3}^d}\ket{L_{-3}^c}-\ket{R_{3}^c}\ket{L_{-3}^d}.\tag{B10}
\end{equation}
By placing two polarizers oriented at diagonal direction to eraser the distinguishable information in polarization degree of freedom, we obtain a four-dimensional Bell basis as
\begin{equation}
\ket{\psi_{1,0}^{3}}\rightarrow(\ket{-3}\ket{3}-\ket{3}\ket{-3}+\ket{-1}\ket{1}-\ket{1}\ket{-1})/2\otimes\ket{c}\ket{d}.\tag{B11}
\end{equation}
Thus, by controlling the incident polarization entanglement and the relative phases between different spatial modes, the Bell states $\ket{\psi_{m,n}^{3}}$ can be prepared.

Finally, by flipping the OAM states of paired photons that are entangled in the form of $\ket{\psi_{m,n}^{3}}$, we are allowed to create the Bell states $\ket{\psi_{m,n}^{4}}$.

To conclude, through the tunable controlling of polarization entanglement and the HOM operation for a Bell state filter, we can exploit their combination to prepare the complete four-dimensional symmetric and anti-symmetric Bell states with current linear optics. The specific transformation operations for  the complete four-dimensional OAM Bell states are shown in Table \ref{tab1}.

\section*{APPENDIX C: Experimental result}
\begin{figure}[htbp]
\centering
\includegraphics[width=0.95\linewidth]{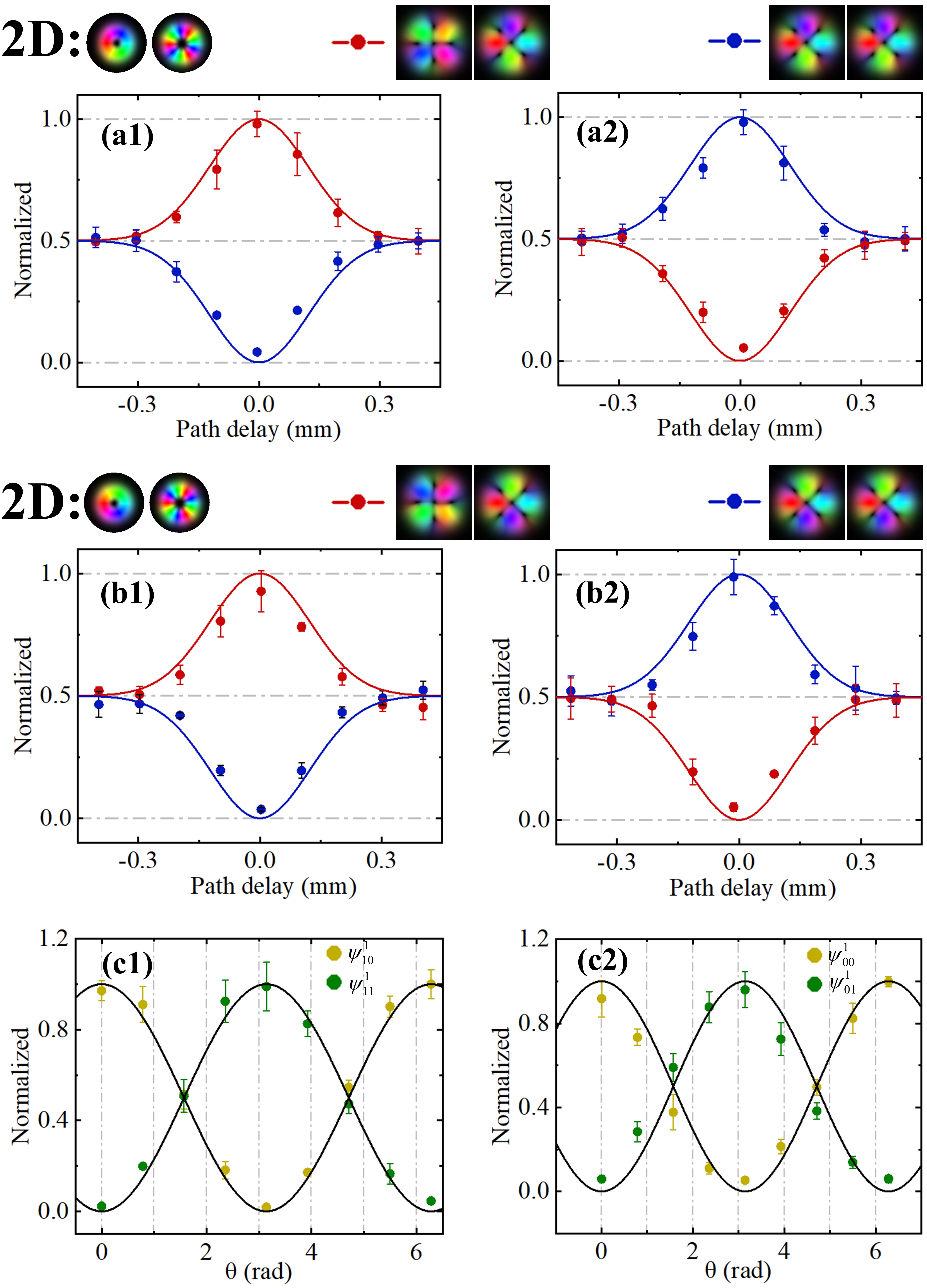}
\vspace{-4mm}
\caption{Experimental observation of HOM interference in two-dimensional subspaces. The circular insets show the input spatial modes of paired photons, and the square insets show their projected measurement bases. (a1-2) HOM interference pattern when the signal photons are projected onto $(\ket{-1}+\ket{3})/\sqrt{2}$, and its partner idler photons are projected onto $(\ket{-1}-\ket{3})/\sqrt{2}$ (red dots) and $(\ket{-1}+\ket{3})/\sqrt{2}$ (blue dots). (b1-2) HOM interference pattern when the signal photons are projected onto $(\ket{1}+\ket{-3})/\sqrt{2}$, and its partner idler photons are projected onto $(\ket{1}-\ket{-3})/\sqrt{2}$ (red dots) and $(\ket{1}+\ket{-3})/\sqrt{2}$ (blue dots). (c1-2) The coincidence detection when the paired photons are projected onto the measurement bases $(\ket{1}+\ket{-1})/\sqrt{2}$ and $(\ket{3}+e^{i\theta}\ket{-3})/\sqrt{2}$. The error bars are standard deviations calculated from multiple consecutive measurements and the curves are calculated from the theoretically expected visibilities.}
\label{figure_6}
\end{figure}
In our experimental implementation, when the input two-photon polarization states are set as $\ket{R}\ket{L}$ and $\ket{L}\ket{R}$ respectively, we can get two different sub-entanglement $(\ket{3}\ket{-1}-\ket{-1}\ket{3})/\sqrt{2}$ and $(\ket{-3}\ket{1}-\ket{1}\ket{-3})/\sqrt{2}$. As shown in Fig. \ref{figure_6}\textcolor{blue}{(a1-b1)}, the peaks are observed if the measurement bases are $(\ket{\mp1}+\ket{\pm3})/\sqrt{2}$ and $(\ket{\mp1}-\ket{\pm3})/\sqrt{2}$, and conversely the dips would be observed if the $(\ket{\mp1}+\ket{\pm3})/\sqrt{2}$ and $(\ket{\mp1}+\ket{\pm3})/\sqrt{2}$. By adding two dove prisms  with relative angles $22.5^{\circ}$ to modulate the relative phase between different spatial modes, the symmetric two-dimensional sub-entanglement $(\ket{3}\ket{-1}+\ket{-1}\ket{3})/\sqrt{2}$ and $(\ket{-3}\ket{1}+\ket{1}\ket{-3})/\sqrt{2}$ are obtained. In this case, the reversed  interference pattern are observed as shown in Fig. \ref{figure_6} \textcolor{blue}{(a2-b2)}. According to the measurement results presented in Fig. \textcolor{blue}{6(a-b)}, the experimentally measured  interference visibility in the two-dimensional Hilbert space is $96.1\pm0.5\%$. If the initial polarization entanglement is $(\ket{R}\ket{L}\pm\ket{L}\ket{R})/\sqrt{2}$, we can observe the cosine and sinusoidal curves as a function of phase parameter $\theta$ that defined in the measurement bases as $(\ket{1}+\ket{-1})/\sqrt{2}$ and $(\ket{3}+e^{i\theta}\ket{-3})/\sqrt{2}$ as shown in Fig. \ref{figure_6}\textcolor{blue}{(c1)}. Analogously, when the initial polarization entanglement is $(\ket{R}\ket{L}\pm i \ket{L}\ket{R})/\sqrt{2}$, the corresponding sinusoidal and cosine curves are shown in Fig. \ref{figure_6}\textcolor{blue}{(c2)}. Finally, the four-dimensional OAM Bell states $\ket{\psi}_{m,n}^{1}$ can be obtained. In addition, when the input two-photon polarization states are set as $\ket{R}\ket{R}$ and $\ket{L}\ket{L}$ respectively, we can obtain  another  two different forms of sub-entanglements $(\ket{3}\ket{1}-\ket{1}\ket{3})/\sqrt{2}$ and $(\ket{-3}\ket{-1}-\ket{-1}\ket{-3})/\sqrt{2}$. Through the tunable control of polarization entanglement and the HOM operation of the Bell state filter, we can prepare four-dimensional Bell states $\ket{\psi}_{m,n}^{2}$. The two-photon interference observations of the relevant two-dimensional subspace are shown in Fig. \ref{figure_2}.

\section*{APPENDIX D: Symmetric property of the complete four-dimensional Bell states}
While the conventional formulation of the complete Bell basis has been proposed, our generated Bell states behave as perfect symmetric and anti-symmetric quantum state. Here, symmetric property of the complete four-dimensional Bell states are shown in Table \ref{tab2}.
\begin{table}[htbp]
\caption{Symmetric property of the complete four-dimensional Bell states.}
\label{tab2}
\begin{ruledtabular}
\begin{tabular}{cccc}
\textrm{State}&
\textrm{Symmetry}&
\textrm{State}&
\textrm{Symmetry}\\
\colrule
$\ket{\psi}_{1,0}^{1}$ & Anti-symmetric & $\ket{\psi}_{1,1}^{2}$ & Anti-symmetric \\
\hline
$\ket{\psi}_{1,1}^{1}$ & Anti-symmetric & $\ket{\psi}_{1,1}^{2}$ & Anti-symmetric \\
\hline
$\ket{\psi}_{0,0}^{1}$ & symmetric & $\ket{\psi}_{0,0}^{2}$ & symmetric \\
\hline
$\ket{\psi}_{0,1}^{1}$ & symmetric & $\ket{\psi}_{0,1}^{2}$ & symmetric \\
\hline
$\ket{\psi}_{1,0}^{3}$ & Anti-symmetric & $\ket{\psi}_{1,0}^{4}$ & symmetric \\
\hline
$\ket{\psi}_{1,1}^{3}$ & Anti-symmetric & $\ket{\psi}_{1,1}^{4}$ & symmetric \\
\hline
$\ket{\psi}_{0,0}^{3}$ & symmetric & $\ket{\psi}_{0,0}^{4}$ & symmetric \\
\hline
$\ket{\psi}_{0,1}^{3}$ & symmetric & $\ket{\psi}_{0,1}^{4}$ & symmetric \\
\end{tabular}
\end{ruledtabular}
\end{table}

\section*{APPENDIX E: Overlap between states}
\begin{table}[htbp]
\caption{Overlap between experimentally measured states and theoretically expected states.}
\label{tab3}
\begin{tabular}{|c|cccccccc|}
\hline
 &$\ket{\psi}_{1,0}^{2}$ & $\ket{\psi}_{1,1}^{2}$ & $\ket{\psi}_{0,0}^{2}$ & $\ket{\psi}_{0,1}^{2}$ & $\ket{\psi}_{1,0}^{1}$ & $\ket{\psi}_{1,1}^{1}$ & $\ket{\psi}_{0,0}^{1}$ & $\ket{\psi}_{0,1}^{1}$\\
\colrule
$\ket{\psi}_{1,0}^{2}$ & 0.7945 & 0.0011 & 0.0015 & 0.0014 & 0.0030 & 0.0006 & 0.0008 & 0.0004\\
$\ket{\psi}_{1,1}^{2}$ & 0.0194	& 0.8047 & 0.0561 & 0.0127 & 0.0011 & 0.0009 & 0.0005 & 0.0006\\
$\ket{\psi}_{0,0}^{2}$ & 0.0008 & 0.0010 & 0.8039 & 0.0015 & 0.0019 & 0.0030 & 0.0005 & 0.0005\\
$\ket{\psi}_{0,1}^{2}$ & 0.0065 & 0.0066 & 0.0079 & 0.8254 & 0.0026 & 0.0033 & 0.0015 & 0.0014\\
$\ket{\psi}_{1,0}^{1}$ & 0.0006 & 0.0009 & 0.0002 & 0.0003 & 0.8065 & 0.0007 & 0.0007 & 0.0009\\
$\ket{\psi}_{1,1}^{1}$ & 0.0005 & 0.0013 & 0.0001 & 0.0001 & 0.0002 & 0.8191 & 0.0004 & 0.0016\\
$\ket{\psi}_{0,0}^{1}$ & 0.0010 & 0.0008 & 0.0013 & 0.0006 & 0.0022 & 0.0039 & 0.7733 & 0.0027\\
$\ket{\psi}_{0,1}^{1}$ & 0.0003 & 0.0005 & 0.0006 & 0.0012 & 0.0069 & 0.0003 & 0.0003 & 0.7973\\
\hline
\end{tabular}
\end{table}
Table \ref{tab3} summarizes the data underlying Fig. \ref{figure_4}, which illustrates the overlap between the experimentally obtained quantum states and the theoretically predicted Bell basis. These experimental outcomes confirm the capability of our method to accurately manipulate the complete set of Bell states. As a result, the generated states are well-suited for practical applications in quantum information processing, such as quantum dense coding and quantum key distribution.

\section*{APPENDIX F: Certification of high-dimensional entanglement}
The dimensionality of entanglement defines the minimum number of levels necessary to accurately characterize a quantum state. For a $d\times d$-dimensional quantum state with a fidelity of $F$ relative to the expected entangled state, the lower bound of its entanglement dimensionality is given by
\begin{equation}
d_\text{ent} = \max\left\{k\,;\,F\leq\frac{k-1}{d}\right\},\tag{F1}
\end{equation}
which has been rigorously established. Consequently, this allows us to estimate the dimensionality of any entangled state based on its fidelity. Applying this inequality, all measured fidelities in our experiment surpass the threshold for a four-dimensional entangled state, confirming that $d_\text{ent} = 4$. This result indicates that the generated quantum state requires a $4 \times 4$-dimensional entangled state for its complete description.

\bibliography{apssamp}

\end{document}